\documentstyle[12pt,titlepage]{article}
\newlength{\extraspace}
\newlength{\extraspaces}
\catcode`\@=11

\def\numberbysection{\@addtoreset{equation}{section}
\def\theequation{\arabic{section}.\arabic{equation}}}

\newcommand{\be}{\begin{equation}
\addtolength{\abovedisplayskip}{\extraspaces}
\addtolength{\belowdisplayskip}{\extraspaces}
\addtolength{\abovedisplayshortskip}{\extraspace}
\addtolength{\belowdisplayshortskip}{\extraspace}}
\newcommand{\ee}{\end{equation}}
\newcommand{\ba}{\begin{eqnarray}
\addtolength{\abovedisplayskip}{\extraspaces}
\addtolength{\belowdisplayskip}{\extraspaces}
\addtolength{\abovedisplayshortskip}{\extraspace}
\addtolength{\belowdisplayshortskip}{\extraspace}}
\newcommand{\ea}{\end{eqnarray}}
\newcommand{\bm}[1]{\mbox{\boldmath ${#1}$}}

\begin{document}
\begin{titlepage}
\addtolength{\baselineskip}{.7mm}
\thispagestyle{empty}
\begin{flushright}
%TIT/HEP-- \\
NUP-A-94-9 \\
hep-th/9406123\\
June, 1994
\end{flushright}
\begin{center}
{\large{\bf SMOOTH PATHS
ON THREE DIMENSIONAL LATTICE}} \\[15mm]
\vspace*{3cm}
%
%{\sc Masako Asano}
%\footnote{\tt e-mail: maa@phys.titech.ac.jp}\\[2mm]
%{\it Department of Physics, Tokyo Institute of Technology, \\[1mm]
%Oh-okayama, Meguro, Tokyo 152, Japan} \\[6mm]
{\sc Chigak Itoi}
\footnote{\tt e-mail: itoi@phys.cst.nihon-u.ac.jp} \\[2mm]
{\it Department of Physics, \\[1mm]
College of Science and Technology, Nihon University, \\[1mm]
Kanda Surugadai, Chiyoda-ku, Tokyo 101, Japan} \\[6mm]
%\ \ and \ \
%{\sc Shin-ichi Kojima}
%\footnote{{\tt e-mail: kotori@phys.titech.ac.jp}, JSPS fellow}\\[2mm]
%{\it Department of Physics,
%Tokyo Institute of Technology, \\[1mm]
%Oh-okayama, Meguro, Tokyo 152, Japan} \\
\vspace*{3cm}
%\vfill
{\bf Abstract}\\[5mm]
{\parbox{13cm}{\hspace{5mm}
A particular class of random walks with
a spin factor on a three dimensional
 cubic lattice is studied.
This three dimensional random walk
model is a simple generalization of
random walk for
the two dimensional Ising model. All critical diffusion
constants and associated critical exponents
are calculated.
Continuum field theories such as Klein-Gordon,
Dirac and massive Chern-Simons theories are
constructed near several critical points.
}}
\end{center}
\vspace*{5ex}
\end{titlepage}
\setcounter{section}{0}
\setcounter{equation}{0}
%\addtolength{\baselineskip}{2mm}
%
%%%%%%%  Introduction  %%%%%%%%%%%%%%%%%%%%%%%%%%%%%%%%%%%%%%%%%%%
%\section{Introduction}
It is well-known that relativistic field theory of
spinning particles
in three dimensional
Euclidian space
can be described in a %naive
continuum random walk representation
\cite{P} - \cite{T}.
The free energy of spinning particles with spin $J$
is represented as a particle path integral
by using SU(2) coherent state \cite{Ch},
and eventually one obtains a geometric form
\be
{\rm Tr} \log[J^{-1} J_\mu \partial _\mu + m ]
= \int \frac{{\cal D}{\bm x}}{V_{Diff}}
 \exp \left[ -m \int dt \sqrt{ \dot{{\bm x}} ^2}
 + i J \Phi [ \frac{\dot{{\bm x}}}{\sqrt{ \dot{{\bm x}} ^2}}
 ] \right],
\label{cr}
\ee
where $J_\mu$ is a spin matrix with
the magnitude $J$ and $\exp [i J \Phi ]$ is the spin factor.
$\Phi[{\bm e}]$ is defined as the area enclosed by
a path ${\bm e}(t)$
on a two dimensional unit sphere.
In the right hand side of (\ref{cr}),
the value of $\Phi$ is a functional of
the unit tangential vector on a path of the particle.
This model shows some nontrivial critical behaviors involved
in different universality class from Brownian motion.
A typical path becomes much smoother than
that in the Brownian motion
because of the spin factor.
Then, this model is sometimes called
a rigid path model.
This representation of spinning particles (\ref{cr}) has been
employed to argue the Bose-Fermi transmutation
in three dimensional relativistic field
theory \cite{P} \cite{IIM} \cite{T}.
However,
the naive continuum theory yield
some subtle problems.
 The point splitting
regularization employed there neglects
contributions of paths with intersection points,
which may affect the critical behavior.
Despite the fractal structure of
paths in random walk, the topological formula used in
the argument is verified by a differential geometry of the path.
Although a typical path of spinning particles is much smoother
than that of the Brownian motion, we can claim only its possibility.
The reliability of this
non-perturbative consequence derived
by a continuuum theory remains unclear.
Thus, more rigorous studies based on lattice
regularization are desired. \\

In this letter, we construct
continuum field theories of relativistic
spinning particles in terms of random walks
on a three dimensional lattice.
Random walk with the spin factor
is defined on lattice space instead of the
naive continuum path integral (\ref{cr}).
Here, the continuum limit in this model is discussed more rigorously.
This random walk model
is a simple generalization of the diffusion process
used for solving the two dimensional Ising model
\cite{V} to three dimensional
models with other spins.
We find all critical points and study
their critical behaviors.
Continuum field theories of
Klein-Gordon, Dirac and massive Chern-Simons
fields can be constructed at corresponding
critical points. These results are consistent with
the continuum path integral (\ref{cr}). \\

%%%%%%%  Definition of the model  %%%%%%%%%%%%%%%%%%%%%%%%%%%%%%%%%%%
%\section{Definition of the model}

We consider a Markovian diffusion process
on a three dimensional cubic lattice
with an amplitude depending on
the preceding step. To obtain the free energy,
we evaluate
a sum over all paths leaving and returning to
the origin. Let ${\bm x}$ and ${\bm e}_\alpha$ be a
lattice site and a bond indices, respectively.
The length of each bond is taken to be unity.
We construct a diffusion process on lattice
which corresponds to the naive continuum
path integral for spinning particles.
We define a probability  amplitude
$A^{\alpha \beta}_{t}({\bm x})$
of finding the particle after $t$ steps
at a site ${\bm x}$
coming from ${\bm x}-{\bm e}_\beta$.
Here, $A^{\alpha \beta}_{t}({\bm x})$
 satisfies a recursion relation
\be
A^{\alpha \beta}_{t+1}({\bm x})
= \sum_{\gamma= \pm 1} ^{\pm 3}
\kappa \ e^{i J S({\bm e}_3, {\bm e}_\beta, {\bm e}_\gamma)}
A^{\alpha \gamma}_t ({\bm x}-{\bm e}_\beta),
\label{rec}
\ee
where $\kappa$ is a diffusion constant,
$J$ is a spin parameter and
$S({\bm e}_\alpha, {\bm e}_\beta, {\bm e}_\gamma)$
is the area of a spherical triangle with vertices
${\bm e}_\alpha$, ${\bm e}_\beta$ and ${\bm e}_\gamma$.
The sum is taken over all nearest neighbor bonds ${\bm e}_\gamma$
at the site ${\bm x}$,
where we denote ${\bm e}_{-\gamma} =-{\bm e}_{\gamma}$.
In our model any back tracking process
is forbidden, that is
\be
A^{\alpha \beta}_t({\bm x}+{\bm e}_\beta)=0.
\ee
The initial condition is given by
\be
A^{\alpha \beta}_0 ({\bm x})
 = \delta^{\alpha \beta}
  \delta_{x_1 0}~ \delta_{x_2 0}~ \delta_{x_3 0}.
\ee
This model is expected to be a lattice version
of the continuum random walk model (\ref{cr}).
Here, we consider only half integer $J$ for simplicity.
This diffusion process
in two dimensions with $J= \frac{1}{2}$ is well-known as
a convinient tool for calculating
the partition function of the two dimensional
Ising model. In this representation,
Majorana spinor is obtained as
a critical theory of the
two dimensional Ising model.
Our three dimensional diffusion model can
be regarded as its generalization.
As in the two dimensional
Ising model, we define a free energy $f(\kappa, J)$
per unit volume by
\be
f(\kappa, J)=(-1)^{2J} \sum_{t=1} ^\infty
\sum_{\alpha= \pm 1} ^{\pm 3} \frac{1}{t}
A^{\alpha \alpha}_t({\bf 0}).
\label{fe1}
\ee
In momentum space we
obtain the Fourier transformed
recursion relation as
a block diagonal form
\be
B^{\alpha \beta}_{t+1}({\bm p})
= \sum_\gamma \kappa B^{\alpha \gamma} _t ({\bm p})
Q^{\gamma \beta} ({\bm p}, J),
\label{frc}
\ee
where $B^{\alpha \beta}_t (\bm p)$ is the Fourier transform
of the amplitude $A^{\alpha \beta} _t (\bm x)$.
The matrix $Q({\bm p}, J)$ has the form
\be
 Q({\bm p}, J)=\left[
\matrix{e^{-ip_2} & e^{ip_1 +i \frac{\pi}{2} J}
& 0 & e^{-ip_1 -i \frac{\pi}{2} J}
& e^{-ip_3} & e^{i p_3 -i \pi J} \cr
e^{-ip_2-i \frac{\pi}{2} J} & e^{i p_1}
& e^{ip_2+i \frac{\pi}{2} J} & 0
& e^{-i p_3} & e^{ip_3 -i2 \pi J} \cr
0 & e^{ip_1 -i \frac{\pi}{2} J}
& e^{ip_2} & e^{-ip_1 +i \frac{\pi}{2} J}
& e^{-ip_3} & e^{ip_3+ i \pi J} \cr
e^{-i p_2+i \frac{\pi}{2} J} & 0
& e^{ip_2-i \frac{\pi}{2} J} & e^{-ip_1}
& e^{-ip_3} & e^{ip_3} \cr
e^{-ip_2} & e^{ip_1}
& e^{ip_2} & e^{-ip_1}
& e^{-ip_3} & 0 \cr
e^{-ip_2 + i \pi J} & e^{i p_1 + i 2 \pi J}
& e^{i p_2 - i \pi J} & e^{-i p_1}
& 0 & e^{i p_3} \cr} \right]
\ee
The solution of the recursion equation (\ref{frc})
is
\be
B_{t}({\bm p}) =
  \{\kappa Q({\bm p}, J) \}^t.
  \ee
 The free energy can be written in terms of
the matrix $Q({\bm p}, J)$
\be
f(\kappa, J)=(-1)^{2J} \int \frac{d^3 p}{(2 \pi)^3}
{\rm Tr} \sum _{t=1} ^{\infty}
\frac{1}{t} \{ \kappa Q({\bm p}, J) \} ^t
\label{sum}
\ee
This sumation is calculated at least formally.
\be
f(\kappa, J)=(-1)^{2J} \int \frac{d^3 p}{(2 \pi)^3}
{\rm Tr} \log[1 - \kappa Q({\bm p}, J)]
\label{fe2}
\ee

%%%%%%%%%%%%%%%%%%%%% Critical Theories %%%%%%%%%%%%%%%%%%%%%%%%%%%%%%%%%%%%%
%\section{Critical Theories}

Physical quantities become nonanalytic
at the critical point as a function of the
diffusion constant $\kappa$.
If the fundamental length scale diverges
at the critical point,
the critical behavior
can be described by
a continuum field theory. Our model
at several critical points can be described by
certain relativistic spinning fields
whose mass matrices are determined by
$Q({\bf 0}, J)$. A critical diffusion constant $\kappa_c$
is determined by the eigenvalue equation of
$Q({\bf 0}, J)$
\be
\det [\kappa_c ^{-1}- Q({\bf 0}, J)] = 0.
\label{ee}
\ee
At $\kappa=\kappa_c$, the free energy becomes
nonanalytic. Since $S({\bm e}_\alpha, {\bm e}_\beta,
{\bm e}_\gamma)$ in (\ref{rec})
is quantized by $\frac{\pi}{2}$ in our idenical
cubic lattice, the eigenvalue equation
possess the symmetry $J \rightarrow 4-J$.
If we consider half-integer $J$,
then $J$ takes
$J=0, \ \frac{1}{2}, \ 1, \ \frac{3}{2}, \ 2$.
For each $J$, we find solutions of
(\ref{ee}) as follows: \\
\ \\
\begin{tabular}{rlllllllll}
$J$ & = &  0, \ & $\kappa_c ^{-1}$ =
& $-$1, & $-$1, & 1, & 1, & 1, & 5,  \\
\ \\
$J$ & = & $\frac{1}{2}$, \ & $\kappa_c ^{-1}$ =
& 1$-\sqrt{2}$, & 1$-\sqrt{2}$, & 1$-\sqrt{2}$, & 1$-\sqrt{2}$,
& $1+ 2 \sqrt{2}$, & $1+ 2 \sqrt{2}$, \\
\ \\
$J$ & = & 1, \ & $\kappa_c ^{-1}$ =
& $-$1, & $-$1, & $-$1, & 3, & 3, & 3, \\
\ \\
$J$ & = & $\frac{3}{2}$, \ & $\kappa_c ^{-1}$ =
& $1-2 \sqrt{2}$, & $1-2 \sqrt{2}$,
& 1+$\sqrt{2}$, & 1+$\sqrt{2}$, & 1+$\sqrt{2}$, & 1+$\sqrt{2}$, \\
\ \\
$J$ & = & 2, \ & $\kappa_c ^{-1}$ =
&  $-$3, & 1, & 1, & 1, & 3, & 3. \\
\end{tabular} \\

One can see several degeneracies.
At $n$-th degenerate critical point,
the amplitude is expressed as a linear combination of
$n$ independent eigenvectors and
can survive as an excitation with long range correlation.
One can trust the formula (\ref{fe2})
only for bigger $|\kappa|^{-1}$ than the maximal absolute
value of eigenvalues
$|\kappa_c |^{-1}$ because of the convergent radius
of the infinite sum (\ref{sum}).
Thus we have to be concerned about only critical points
$(J, \kappa_c ^{-1})=(0,5), (\frac{1}{2}, 1+\sqrt{2}),
(1,3), (\frac{3}{2}, 1+\sqrt{2})$, (2, -3) and (2, 3).
One can constructs
relativistic field theories at critical points
$(J, \kappa^{-1})=(0, 5)$, $(\frac{1}{2}, 1+ 2 \sqrt{2})$,
$(1, 3)$ and $(2, -3)$, if one tunes the diffusion constant
$\kappa$ to these critical points $\kappa_c$.
Those with single, double and triple
degeneracies give us Klein-Gordon, Dirac and
massive Chern-Simons fields, respectively.
We do not find any other critical points that
give continuum field theory. \\

First, we show the behavior of the free energy at
critical points which give relativistic field theories.
At $(J, \kappa^{-1})=(0, 5)$ and $(2, -3)$,
the critial behaviors belong to the same
universality class. The free energy at $(0, 5(1+s))$ is
given
\be
f(\kappa, 0) \simeq \int \frac{d^3 p}{(2 \pi)^3}
\log[{\bm p}^2+4s + O(p^4)], % \\
%f(\kappa, 2) \simeq \int \frac{d^3 p}{(2 \pi)^3}
% \log[{\bm p}^2+12s ].
\label{k}
\ee
where we introduce small parameter $s$
$$
s = \left| \frac{\kappa-\kappa_c}{\kappa_c} \right|.
$$
This critical behavior belongs to the same universality class
of the Brownian motion. The correlation length diverges at
the critical point as $s^{-\frac{1}{2}}$, thus
the critical exponent $ \nu = \frac{1}{2}$.
This critical point gives the Klein-Gordon theory as the
continuum limit.
The free energy near these two critical points behave as \\
$$
f(\kappa) \sim s^{\frac{3}{2}} \log s.
$$
This indicates that the order of the phase transition
is second and the critical
exponent $\alpha= \frac{1}{2}$. \\

At $(J, \kappa^{-1})=(\frac{1}{2}, 1+ 2 \sqrt{2})$ ,
the free energy is
\be
f(\kappa,{\textstyle \frac{1}{2}})
\simeq - \int \frac{d^3 p}{(2 \pi)^3}
\log[{\bm p}^2 + s^2 + O(p^4)] \\
\sim s^{3} \log s.
\label{d1}
\ee
The correlation length becomes $s^{-1}$, and then
$\nu=1$ and $\alpha=-1$.
The two component Dirac field is obtained
as the continuum limit at this critical point.
\be
f(\kappa, {\textstyle \frac{1}{2}})
\simeq - \int \frac{d^3 p}{(2 \pi)^3}
{\rm Tr} \log[- i {\bm p} \cdot {\bm \sigma} -s],
\label{d2}
\ee
where ${\bm \sigma}$ is the Pauli matrix
used as the gamma matrix in
three dimensions.
Other four components
correspond to heavy particles which does not affect
critical phenomena. \\

At $(J, {\kappa}^{-1}) = $  (1, 3),
the nonanalytic part of the free energy at
$\kappa= 1+ \frac{s}{3}$ is
evaluated as follows:
\be
f(\kappa, -1) \simeq \int \frac{d^3 p}{(2 \pi)^3}
 \log[\frac{16}{27} s {\bm p}^2 +
 \frac{64}{27} s^3 + O(p^4)]  \\
\sim s^{3} \log s.
\label{c1}
\ee
The free energy can be written also in terms of
the following masssive vector field
\be
f(\kappa, -1) \simeq \int \frac{d^3 p}{(2 \pi)^3}
{\rm Tr} \log[i {\bm p} \cdot {\bm L} + 4 s],
\label{c2}
\ee
where the matrix {\bm L} is a spin matrix with the
magnitude 1.  One can take
$(L_\mu)_{\nu \lambda} = i \epsilon_{\mu \nu \lambda}$.
Two critical exponents
$\alpha=-1$ and $\nu=1$ are same as those of the Dirac field.
The continuum limit is taken practically by
$s \rightarrow 0$ after the replacement
${\bm p} \rightarrow s^{\nu} {\bm p} $.
The model has infinite ultraviolet cutoff by this procedure.
The lagrangians of three types of relativistic
field theories constructed above
are given for the Klein-Gordon field
\be
{\cal L}_0 = \partial _\mu \phi ^{\ast} \partial _\mu \phi
+ m \phi^{\ast} \phi,
\ee
the Dirac field
\be
{\cal L}_{\frac{1}{2}} = \bar{\psi} (-i \gamma _\mu \partial _\mu
+ m) \psi,
\label{ld}
\ee
and the massive U(1) Chern-Simons field
\be
{\cal L}_1= -i \epsilon ^{\mu \nu \lambda}
A_\mu ^{\ast} \partial_\nu A_\lambda
 + m A_\mu ^{\ast} A_\mu.
\label{lv}
\ee
Above these three types of critical theories agree with
the result given by the naive continuum path integral (\ref{cr}). \\

The order of phase transition
at $(J, \kappa^{-1})=
(\frac{3}{2}, 1+\sqrt{2})$, and $(2, 3)$
becomes higher than the second and then
the continuum limit can exist at those critical points.
One might expect spin $\frac{3}{2}$ relativistic
field theory at these critical points.
Nevertheless, these continuum theories
possess no rotational symmetry, and thus these
obtained continuuum theories cannot
be relativistic. For example, the
free energy at $(J, \kappa^{-1})=(\frac{3}{2},
(1+\sqrt{2})(1+s))$ becomes
\be
f(\kappa, {\textstyle \frac{3}{2}})
\simeq - \int \frac{d^3 p}{(2 \pi)^3}
\log[18(3-2 \sqrt{2}) \{ s^2 {\bm p}^2 + \frac{s^4}{2} + \frac{1}{4}
( p_2 ^2 p_3 ^2 + p_3 ^2 p_1 ^2 + p_1 ^2 p_2 ^2) + O(p^6) \}].
\label{rs}
\ee
This shows no rotational invariance,
even though one takes the limit $s \rightarrow 0$. \\
Even if one neglects the rotational non-invariant term
of $O(p^4)$ in (\ref{rs}), the result does not agree with that of
the naive continuum path integral (\ref{cr}).
Fourth degenerate critical point is supposed to
correspond to $J=\frac{3}{2}$ continuum
path integral (\ref{cr}) which gives
\be
{\rm Tr} \log[ - {\partial}^2 + m^2 ]
+ {\rm Tr} \log[- \frac{1}{9} {\partial}^2 + m^2 ].
\ee
This differs from (\ref{rs}) even in the long wave approximation.
%This indicates that the continuum path integral
%with $J=\frac{3}{2}$ contains twice degrees of freedom as
%many as the lattice random walk model.
%The eigenvalues of four by four matrices in
%the linearized kinetic term of the lattice model
%are $-1$, 0, 0 and 1.
%This fact implies that they are not spin matrices unlike
%spin $\frac{1}{2}$ and 1 cases.

%%%%%%%%%%%%%%%%%%%%%% Conclusion %%%%%%%%%%%%%%%%%%%%%%%%%%%%%%%%%%%%%%%%%
%\section{Conclusion}
In this letter, we studied quantum diffusion process
with a spin factor. Relativistic field theories of Klein-Gordon,
Dirac and massive Chern-Simons fields have been
constructed by taking continuum limit
near critical points. On the other hand, however, field
theory with a spin higher than 1
have not been obtained.
It remains to check the universality of
these obtained results in other lattice spacing.
More detailed and subsequent
works are going to be reported soon. \\

The author would like to thank M. Asano and
S. Kojima for helpful discussions. He is grateful
also to T. Hara for critical comments. He
thanks to T. Fujita for careful reading the manuscript.

\end{document}